\documentstyle[11pt,newpasp,twoside,psfig]{article}
\def\edcomment#1{\iffalse\marginpar{\raggedright\sl#1\/}\else\relax\fi}
\reversemarginpar

\begin{document}
\title{G343.1-2.3 and PSR 1706-44}
\author{R Dodson, M Howlett}
\affil{School of Maths \& Physics, University of Tasmania, Australia}
\author{K. Golap}
\affil{National Radio Astronomy Observatory, Socorro, USA}
\author{N. UdayaShankar}
\affil{RRI, Bangalore, India}
\author{J. Osborne}
\affil{University of Durham, South Rd, Durham, UK}

\begin{abstract}
The association of G343.1-2.3 and PSR 1706-44 has been controversial
from its first proposal. In this paper we present new evidence from
images made by with the Australia Telescope Compact Array (ATCA), MRT
and Mt. Pleasant. To cover the full extent of G343.1-2.3 with ATCA
mosaicing was required, and we present the polarisation images from
this experiment. The ATCA observations confirms the much larger extent
of the SNR, which now encompasses the pulsar.
\end{abstract}
\section{Introduction}

The discovery that the {\it COS-B} gamma-ray source was a pulsar (PSR
B1706-44) was made by Johnston et al (1992). McAdam, Osborne, \& Parkinson
(1993) published a map made by the MOST telescope at 843~MHz of the
area around the gamma-ray source.  It showed a semicircular arc of
emission, which has subsequently been denoted as the supernova remnant
(SNR) G343.1-2.3, with the pulsar seemingly embedded in a small
feature at its south eastern extremity. It was argued that the
approximate distance of 3~kpc for the remnant, derived from the
surface brightness-diameter relationship ($\Sigma$-D) was not
incompatible with the pulsar's dispersion measure, although the
distance indicated by the widely used interstellar electron density
model of Taylor \& Cordes (1993) would be only 1.8~kpc. The $\Sigma$-D
relationship was based on the flux values from single dish
observations, as both the VLA and the MOST integrated flux fell short
of what would be expected from these (McAdam et al 1993).

If the association of the pulsar and the supernova remnant were real a
transverse velocity of $\sim900$~kms$^{-1}$ would be required for the
pulsar to have moved from the approximate geometric centre of the SNR
arc in the characteristic spin down time of 17~kyr. This is a high
value but not the highest measured value for a pulsar nor as high as
that implied for some seemingly well-established pulsar-SNR
associations.

Frail, Goss, \& Whiteoak (1994) mapped the area around the pulsar at
20~cm and 90~cm with the VLA and cast some doubt on it's association
with G343.1-2.3 based on the morphology, and the argument that the
dispersion measure and $\Sigma$-D distance were discrepant. The latter
was subsequently countered by a 21~cm hydrogen line absorption
measurement of the pulsar by Koribalski et~al. (1995) which gives a
distance range of 2.4 to 3.2~kpc.

Nicastro, Johnston, \& Koribalski (1996) also argue against the
association. They report a maximum value only 27~kms$^{-1}$ for the
magnitude of the transverse velocity of the pulsar calculated from the
interstellar scintillation. However in both Dodson (1997) and then
their later paper (Johnston, Nicastro, \& Koribalski 1998) this
estimate is revised up to 100~kms$^{-1}$. The possibility of a direct
determination of the proper motion of PSR B1706-44 has been
investigated, but no phase reference sufficiently strong for the
current Australian VLBI network (the LBA) could be found.

We have made a number of radio images of the remnant to determine its
full extent. These are with MRT at 150~MHz, with the ATCA at 1400~MHz
and at 6.6~GHz with the 26m dish at Mt Pleasant observatory, Hobart.

\section{Images from MRT and Mt. Pleasant}

\begin{figure}
 \hspace{3cm}
 \psfig{file=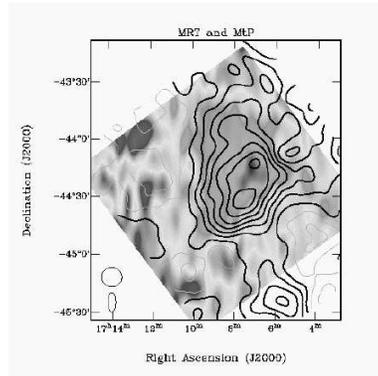,width=5cm}
 \caption{Low resolution, but complete, image from MRT overlaid with
 contours from Mt. Pleasant}
 \label{fig:mrt+mtp}
 \end{figure}

The object as seen at 150~MHz, and at 6.6~GHz are shown in figure
\ref{fig:mrt+mtp}. The 150-MHz map is a preliminary result from the
MRT all sky survey. The 6.6-GHz image was produced as part of the
Honours thesis of M. Howlett (Howlett 1998), a survey along the
galactic plane using the 26m antenna run by the University of
Tasmania. Both contain all spatial frequencies up to their
resolution. They both show broad extended structure to the south of
the bright arc seen by MOST.

\section{Results from the Compact Array}

The Australia Telescope Compact Array has been use to re-observe
G343.1-2.3, with a nineteen pointing mosaic. Mosaicing was required to
cover the extent of the source, but also to assist in the recovery of
the flux on the shortest {\em uv} spacings, which were missed in the
MOST and the VLA observations.  

\begin{figure}
% \begin{center}
 \psfig{file=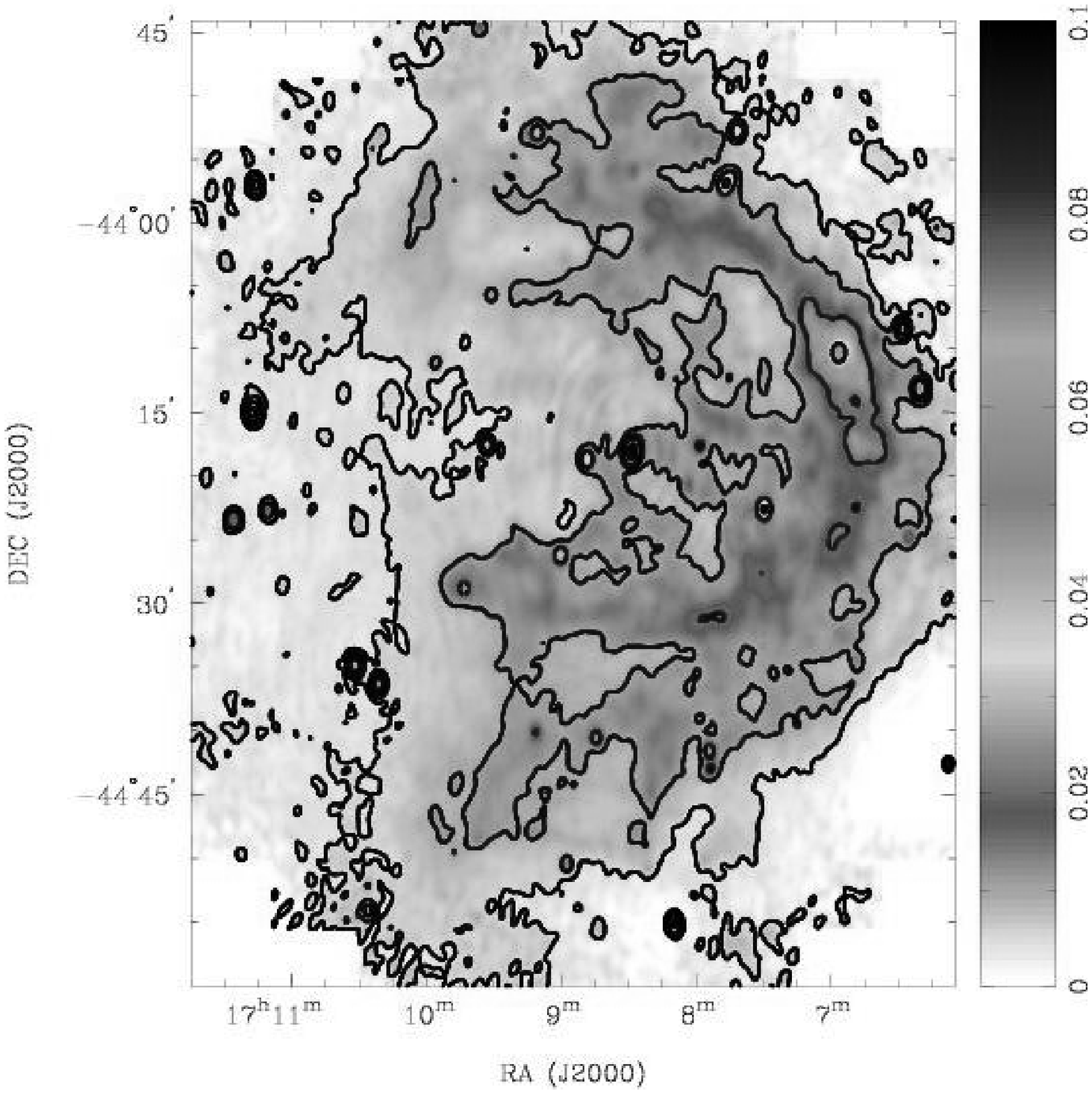,width=5cm}
 \vspace{-4.8cm} \hspace{7cm} 
 \psfig{file=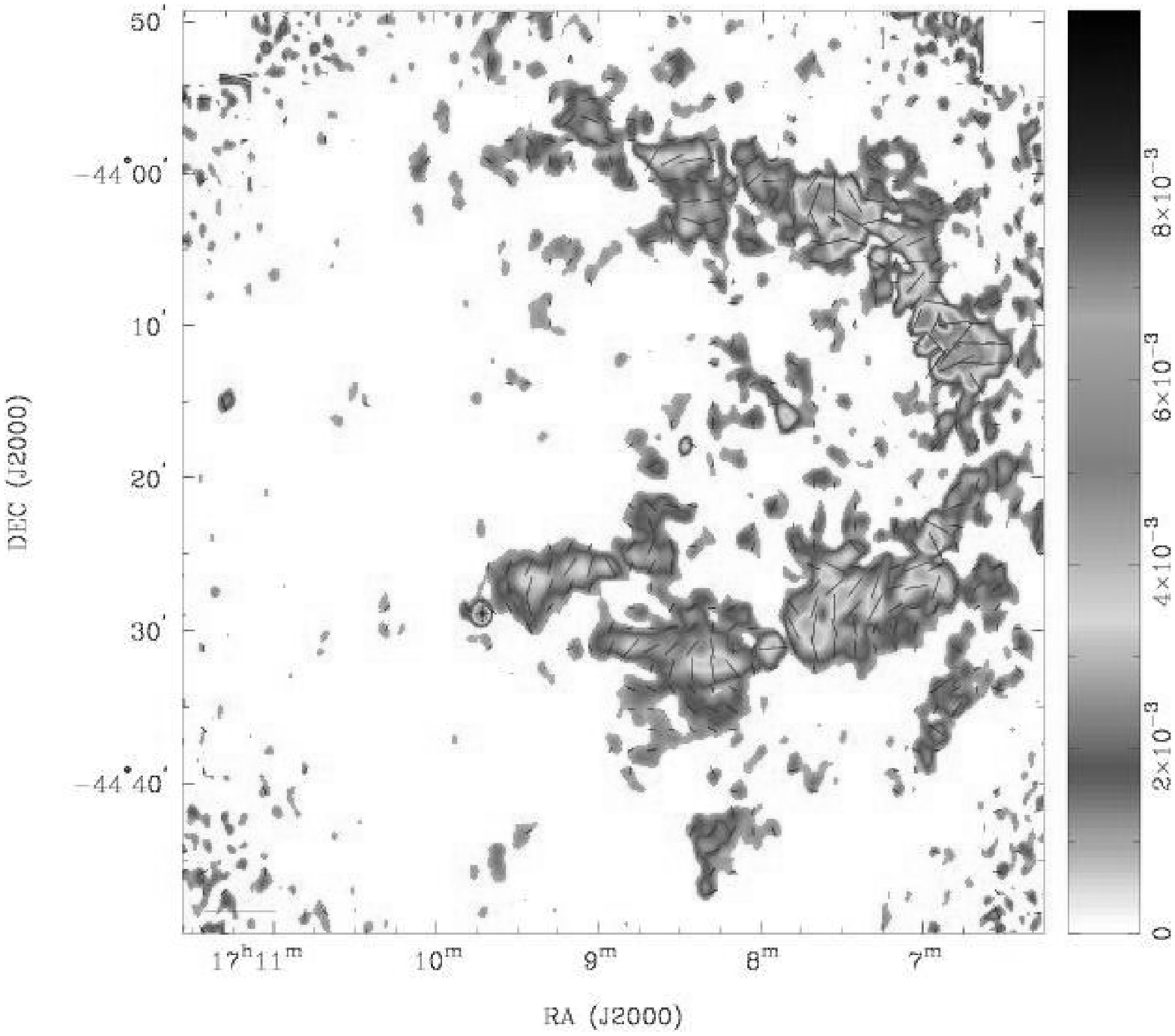,width=5cm,angle=270}
 \vspace{1cm}
 \caption{a) Supernova Remnant G343.1-2.3 at 1384 MHz, and b)
 G323.1-2.3 polarised intensity, with polarisation angle overlaid} 
 \label{fig:at_i}
% \end{center}
 \end{figure}

The images in Figure \ref{fig:at_i} a), look very much like those of MOST
and indeed have very similar resolution ($70^{\prime
\prime} \times 47^{\prime \prime}$) (the longer baselines, to the 6km
antenna, were not used due to their poor support). We have included as
a constraint in the deconvolution the observed total flux for the SNR,
30 Jy, as seen with the 26m single dish in Hobart.

Figure \ref{fig:at_i} b) shows the polarisation angle overlaid on the
polarised intensity. The polarised fraction is typically about 20 per
cent and the polarisation angle is randomly orientated and changes
rapidly around the bright ring. Wisps of polarised emission can be
seen to extend beyond the ring as originally defined, and throughout
the region with broadscale emission. This confirms that the broadscale
emission is associated with the synchrotron emission from the SNR.

It is instructive to compare the VLA maps of figure 4 in Frail et al
(1994) with the map in Figure \ref{fig:at_i} a). The spur of emission
on which the pulsar lies is clearly a continuous feature covering more
than $30'$. This makes it extremely unlikely to be a cometary tail
from the pulsar. Furthermore the width of the spur, $4'$, is unlikely
to be that due to ram pressure of a moving pulsar. It is intriguing
this approximately the size that would be expected from a static
pulsar wind, i.e. the Sedov equations with a continuous input of
energy, with the $\dot{E}$ of PSR1706-44 (Rees \& Gunn 1974).

\section{Conclusion}

The arguments against the association were based around Kaspi's critia
(Kaspi 1996). These were that the pulsar DM distance and the SNR
$\Sigma$-D distance did not match, and the pulsar was at the edge of
the SNR and had too low a velocity to have travelled from the
geometric centre of the SNR. Also there was no morphological evidence
to imply a velocity direction. It has turned out that it is the DM
distance is incorrect, and the extent of the SNR has been
underestimated.

It is clear from both the low total flux values found by the MOST and
VLA maps that there is a smooth broad component to this SNR. The ATCA
observations confirm the extent of the broad emission, and that this
is syncrotron in character. Taking this smooth component as the true
limits of the SNR enlarges the extent and moves it southwards. This
places the pulsar PSR1706-44 within the SNR shell, rather than on the
rim.

We expect a SNR to be associated with PSR1706-44 because of its
youth. The association was rejected in the past, and we have shown
that the grounds for the rejection are not compelling as previously
believed. The question of the pulsar velocity could be resolved by
finding a phase reference for VLBI observation, with either the SKA or
an upgrade to the current LBA. We will be able to rule out the upper
velocity limits with the ATCA within the next few years but will not
be able to measure to the lower limits for a few decades.

\end{document}